# Simulation of cellular irradiation with the CENBG microbeam line using GEANT4

S. Incerti[1], Ph. Barberet, R. Villeneuve, P. Aguer, E. Gontier,
C. Michelet-Habchi, Ph. Moretto, D. T. Nguyen, T. Pouthier, R. W. Smith

*Abstract* -- **Light-ion microbeams provide a unique opportunity to irradiate biological samples at the cellular level and to investigate radiobiological effects at low doses of high LET ionising radiation. Since 1998 a single-ion irradiation facility has been developed on the focused horizontal microbeam line of the CENBG 3.5 MV Van de Graaff accelerator. This setup delivers in air single protons and alpha particles of a few MeV onto cultured cells, with a spatial resolution of a few microns, allowing subcellular targeting. In this paper, we present results from the use of the GEANT4 toolkit to simulate cellular irradiation with the CENBG microbeam line, from the entrance to the microprobe up to the cellular medium.**

## I. Introduction

THE object oriented simulation toolkit GEANT4 allows us to follow ion diffusion through the CENBG microbeam line elements (beam pipe residual gas, collimators, focusing magnetic quadrupoles, single ion transmission detector, exit window, air gap, irradiation well, …) which can increase the spatial and energy dispersions of the beam and degrade the targeting resolution [1]-[3]. First estimations of GEANT4 simulation capabilities at the micrometer scale appear promising, encouraging us to study for the first time the entire experimental setup, paying particular attention to the modelling of the magnetic lenses [4].

## II. The Microbeam Line Setup

The microbeam line of CENBG allows single ion irradiation of individual cells with proton or alpha particles [2], [3]. The incident beam is strongly collimated using a 5 µm circular object collimator and a 10 µm circular diaphragm, 6 m away from the collimator, allowing to reach a low flux mode required for single ion irradiation (typically a few hundred particles on target per second). The residual air pressure of the whole beam line is kept under $5.10^{-6}$ mbar. The beam is then focused using four magnetic quadrupoles in the so called Dymnikov magnetic configuration, leading to symmetrical transverse demagnification factors $D_x = D_y = -1/10$ on target. Their gradients $G_1$ and $G_2$ have been adjusted to focus the beam on the cellular target location, 235 mm away from the physical exit of the last quadrupole. Single ions are counted using a 3.5 mm long isobutane proportional counter running at a gas pressure of 10 mbar ; a 10 µm circular collimator ensures the pressure transition between the beam pipe and the counter. The beam is extracted back into the air through a 150 nm square $Si_3N_4$ window (1 mm²) and is sent through a 100 µm ambient air gap to a 4 µm polypropylen foil, where the HaCat cells have attached and grown. The cells are kept alive in a dedicated well containing Keratinocyte Growing Medium, sealed by a microscope glass slide. The whole line geometry and list of materials have been computed in GEANT4 version 5.2 and are shown on Fig. 1. We have used the low energy electromagnetic package G4LOWEM2.2 and the electronic stopping power table ICRU_R49He. To ensure reliable multiple scattering modelling and to reproduce experimental beam straggling measurements performed on the microbeam line, we have forced an elementary step in each volume equal to one tenth of the corresponding volume size along the beam propagation axis $z$ and the value of the secondary particle cut has been uniquely set to 100 µm [4].

## III. Incident Beam Properties

The modelling of ion beam transportation in the microbeam line requires the knowledge of several beam-optical parameters. In the absence of acceleration along the $z$ propagation axis, the transverse motion of the beam can be represented by ellipses in the phase spaces $(x,x')$ and $(y,y')$ where $x' = p_x / p_z$, $y' = p_y / p_z$ represent respectively the beam angular divergences $\theta$ and $\phi$, and $p_x$, $p_y$, $p_z$ stand for the three coordinates of the beam momentum. The beam is delivered to the microbeam line through a 5 µm diameter circular collimator with a maximum divergence of 0.5 mrad. To quantify higher-order aberrations of the focusing system, we have chosen to describe the beam external envelope at the collimator $z$ location by ellipses, identically on both transverse axes :

$$\frac{x^2\ (\mu m^2)}{2.5^2} + \frac{\theta^2\ (mrad^2)}{0.5^2} = 1, \quad \frac{y^2\ (\mu m^2)}{2.5^2} + \frac{\phi^2\ (mrad^2)}{0.5^2} = 1$$

In a non-accelerating field, the emittance $E$ remains constant and can be estimated from the beam phase space surface $A$ :

$$E = A/\pi = 1.25\ \mu m \times mrad$$

The irradiation setup described in this paper has been optimized for high LET 3 MeV alpha particles. The beam kinetic energy distribution is assumed to be Gaussian with a mean of 3 MeV and a FWHM of 0.150 keV. The corresponding quadrupole focusing gradients have been calculated iteratively with GEANT4 : $G_1 = 3.406 \pm 0.001$ Tm$^{-1}$ and $G_2 = 8.505 \pm 0.001$ Tm$^{-1}$.

---
[1] Corresponding author : Sébastien Incerti
Centre d'Etudes Nucléaires de Bordeaux-Gradignan, IN2P3/CNRS,
Université Bordeaux 1, 33175 Gradignan Cedex, France
Phone : +33 5 57 12 08 89 – Fax : +33 5 57 12 08 01
E-mail : incerti@cenbg.in2p3.fr

## IV. THE FOCUSING QUADRUPLET

### A. Fringing field description

GEANT4 can track a charged particle in any type of magnetic field, as long as the field can be described analytically. It takes approximately 3 s on a Intel Xeon 2.7 GHz PC to track a single ray with a maximum step length of 100 µm in the magnetic field region. In order to calculate the particle trajectories through the quadruplet system, we have chosen to describe the magnetic field profile using the Enge model [5], including the modelling of fringing fields. In the case of a perfect quadrupole, without fringing field, the magnetic field inside the quadrupole is simply given by $B_x = yG$ and $B_y = xG$. When including the fringing field, these expressions become :

$$B_x = y \left[ G - \frac{1}{12}(3x^2 + y^2)\frac{d^2 G}{dz^2} \right] + \text{higher orders}$$

$$B_y = x \left[ G - \frac{1}{12}(3y^2 + x^2)\frac{d^2 G}{dz^2} \right] + \text{higher orders}$$

$$B_z = xy \left[ \frac{dG}{dz} - \frac{1}{12}(x^2 + y^2)\frac{d^3 G}{dz^3} \right] + \text{higher orders}$$

where $G \equiv G(z) = G_0 K(z)$. $G_0$ is the gradient value in the case of a perfect quadrupole. $K(z)$ can be estimated using the experimental profile of the field measured through the lens axis, at a given non-zero radius. Enge uses the following formula :

$$K(z) = \frac{1 + e^{c_0}}{1 + e^{P_n(z)}}$$

where :
$s = (z - z_1)/a_0$ if $z > z_1$
$s = -(z + z_1)/a_0$ if $z < -z_1$
$P_n(z) = c_0 + c_1 s + c_2 s^2$

$z_1$ is the quadrupole positive lower limit of the fringing field region and $a_0$ is the bore radius of the element : $a_0 = 10$ mm. Then,

$$\frac{dK}{dz} = -(1 + e^{c_0}) \frac{dP}{dz} \frac{1}{(1 + e^{P(z)})^2} e^{P(z)}$$

$$\frac{d^2 K}{dz^2} = -(1 + e^{c_0}) e^{P(z)} \times$$

$$\left[ \frac{d^2 P}{dz^2} \frac{1}{(1 + e^{P(z)})^2} + 2 \frac{dP}{dz} \frac{1}{1 + e^{P(z)}} \frac{1}{1 + e^{c_0}} \frac{dK}{dz} \right. $$

$$\left. + \left(\frac{dP}{dz}\right)^2 \frac{1}{(1 + e^{P(z)})^2} \right]$$

$$\frac{d^3 K}{dz^3} = -(1 + e^{c_0}) e^{P(z)} \times$$

$$\left[ \frac{1}{(1 + e^{P(z)})^2} \left( \frac{d^3 P}{dz^3} + 3\frac{d^2 P}{dz^2}\frac{dP}{dz} + \left(\frac{dP}{dz}\right)^3 \right) \right.$$

$$+ 4 \frac{1}{1 + e^{P(z)}} \frac{1}{1 + e^{c_0}} \frac{dK}{dz}\left( \left(\frac{dP}{dz}\right)^2 + \frac{d^2 P}{dz^2} \right)$$

$$\left. + 2 \frac{dP}{dz}\left( \frac{1}{(1+e^{c_0})^2}\left(\frac{dK}{dz}\right)^2 + \frac{1}{1+e^{P(z)}}\frac{1}{1+e^{c_0}}\frac{d^2 K}{dz^2} \right) \right]$$

For a given quadrupole, the uniform field region extends from $z = -z_1$ to $z = z_1$. The fringing field region extends from $-z_2$ to $-z_1$ and from $z_1$ to $z_2$. Beyond $z_2$ (or $-z_2$), the field is zero. The $c_i$ coefficients and the value of $z_1$ must be adjusted in order to fit to the experimental profile of the field, which has not been measured yet in our system. However, we have chosen typical values [6] : $c_0 = -5$, $c_1 = 2.5$, $c_2 = -0.1$, $z_1 = 6$ cm and $z_2 = 13$ cm, leading to an effective length $l_e$ of 16.5 cm (the quadrupole geometrical length is 15 cm and it is distant from the next one by 4 cm), where $l_e$ is defined by :

$$B_0 l_e = \int_{x=y=3 \text{ mm, } z=-z_2}^{x=y=3 \text{ mm, } z=z_2} B_r(z) dz$$

$B_0$ is the value of the field at $z = 0$ and $r = 3\sqrt{2}$ mm is chosen to have a non-zero value of $B_r$. The whole field profile within the quadruplet is shown in Fig. 2.

### B. Intrinsic aberrations

For the chosen gradient configuration, GEANT4 allows the extraction of intrinsic aberration coefficients up to any order from the dependence of the beam transverse position on target as a function of :
- the initial angles $\theta$ and $\phi$, for the spherical aberrations (refer to Fig. 4) :
  $\langle x|\theta \rangle = 1.6$ µm/mrad (astigmatism, first order)
  $\langle y|\phi \rangle = 2.8$ µm/mrad (astigmatism, first order)
  $\langle x|\theta^3 \rangle = -8.7$ µm/mrad$^3$ (spherical, third order)
  $\langle y|\phi^3 \rangle = -25.5$ µm/mrad$^3$ (spherical, third order)
  $\langle x|\theta\phi^2 \rangle = -39.3$ µm/mrad$^3$ (spherical, third order)
  $\langle y|\phi\theta^2 \rangle = -38.7$ µm/mrad$^3$ (spherical, third order)
- the initial beam transverse position and $\delta = \Delta p / p$ for the second order chromatic aberrations :
  $\langle x|x\delta \rangle = -0.02$ µm/mrad%
  $\langle y|y\delta \rangle = -0.02$ µm/mrad%
  $\langle x|\theta\delta \rangle = -103.7$ µm/mrad%
  $\langle y|\phi\delta \rangle = -153.2$ µm/mrad%

From these coefficients, it is possible to illustrate the contribution of the high order aberrations to the beam spot shape on target, as shown in Fig. 5, allowing the design of specific collimator geometries to remove the image distortions. A precise determination of these coefficients could also help in the iterative optimization of the quadruplet field gradients.

### C. Tracking precision

The study of the transverse displacement in the image plane as a function of shooting angles over a wide range of angles ($10^{-1}$ mrad down to $10^{-10}$ mrad) shows a smooth polynomial variation in both planes (refer to Fig. 6) and gives us confidence in the GEANT4 tracking capabilities at this scale.

## V. CELLULAR IRRADIATION

### A. Beam energy and spatial distributions on target

The beam energy and spatial distributions obtained with 20000 incident alphas are shown in Fig. 7 for a pipe residual air pressure of $5.10^{-6}$ mbar. A Gaussian fit to the energy distribution gives us the following estimate : $\langle T \rangle \pm \sigma_T = 2.37 \pm 0.01 \text{ MeV}$. The alpha beam has consequently lost $633 \pm 13$ keV before reaching the target cell. When the beam pipe residual air is replaced by vacuum, the energy loss does not change significantly.

### B. Targeting probability

From these distributions, we can estimate the probability of targeting an alpha particle in a given area at the target location, typically a 10 µm diameter circular area. Assume $N_d$ is the number of alphas detected by the isobutane counter and $N_a$ is the number of particles among them which spread at the target location inside the 10 µm diameter area, then the probability $p_a$ is simply defined as $p_a \pm \sigma_{p_a} = N_a / N_d \pm \sqrt{p_a(1-p_a)/N_d}$ [7]. In the case of a pure vacuum beam pipe, the probability reaches $p_a \pm \sigma_{p_a} = (99.4 \pm 0.1)\%$ and decreases down to $p_a \pm \sigma_{p_a} = (70.5 \pm 0.8)\%$ when the pressure is raised to $5.10^{-6}$ mbar. Experimentally, $p_a$ can be estimated by replacing the irradiation well by a PIN diode with a 10 µm entrance collimator. The diode counts the alpha particles spread on the collimator aperture. We have measured $p_a \approx 80-90\%$ at $5.10^{-6}$ mbar, which is in reasonable agreement with our simulation.

### C. Dose calculation

We have estimated the dose deposited by the alpha beam in a typical HaCat cell, whose cytoplasm and nucleus have been modelled from confocal microscopy images as tubes of elliptical cross section fixed on the polypropylen foil along their revolution axis. For the cytoplasm, the half axes are 4.35 µm and 7.3 µm long and the tube length is 15 µm. For the nucleus, the half axes are 3.5 µm and 6.25 µm long and the tube length is 9.5 µm. Both are made of water. The corresponding dose distributions are shown in Fig. 8. The dose distribution in the cytoplasm shows two populations : the low dose part corresponding to alpha particles that have crossed both the cytoplasm and the nucleus, and the higher dose part, corresponding to alphas that have hit the cytoplasm near the edge, without reaching the nucleus. The most probable dose deposit in the nucleus reaches :

$$D_n \approx 0.33 \text{ Gy}$$

From this estimation, it appears that a 3 MeV alpha beam may deliver doses to cellular nuclei up to a few tenths of a Gray.

## VI. CONCLUSIONS

This study shows GEANT4's capabilities and flexibility in the simulation of cellular irradiation experimental setups at the micrometer scale. The CENBG irradiation microbeam line will soon provide experimental data at this scale that will contribute to validate our simulations, especially in the framework of the new GEANT DNA project. In the near future (2005), a new generation Singletron accelerator will be installed at CENBG and a nanobeam line will be developed for cellular irradiation at the nanometer scale. We believe GEANT4 and its extensions will become the state of the art in the high precision simulation of light ion irradiation setups at the micrometer and nanometer scales.

## VII. ACKNOWLEDGMENTS

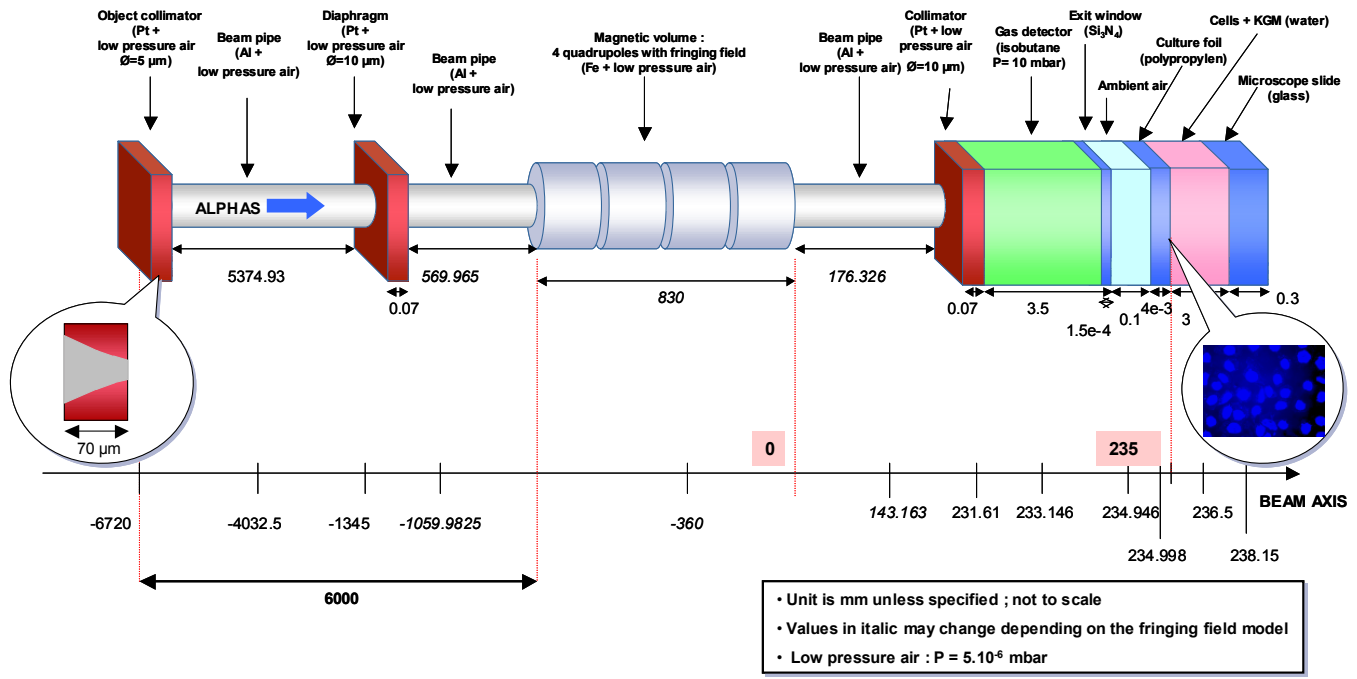

Fig. 1. Details of the microbeam line geometry and materials along the $z$ propagation axis, as they are defined in GEANT4. The cultured cells are located 235 mm away from the physical exit of the last quadrupole. Note the modelling of the collimators' geometry ; they are defined as an assembly of two joined conical sections to reproduce the shape observed by the mean of an electronic microscope [4].

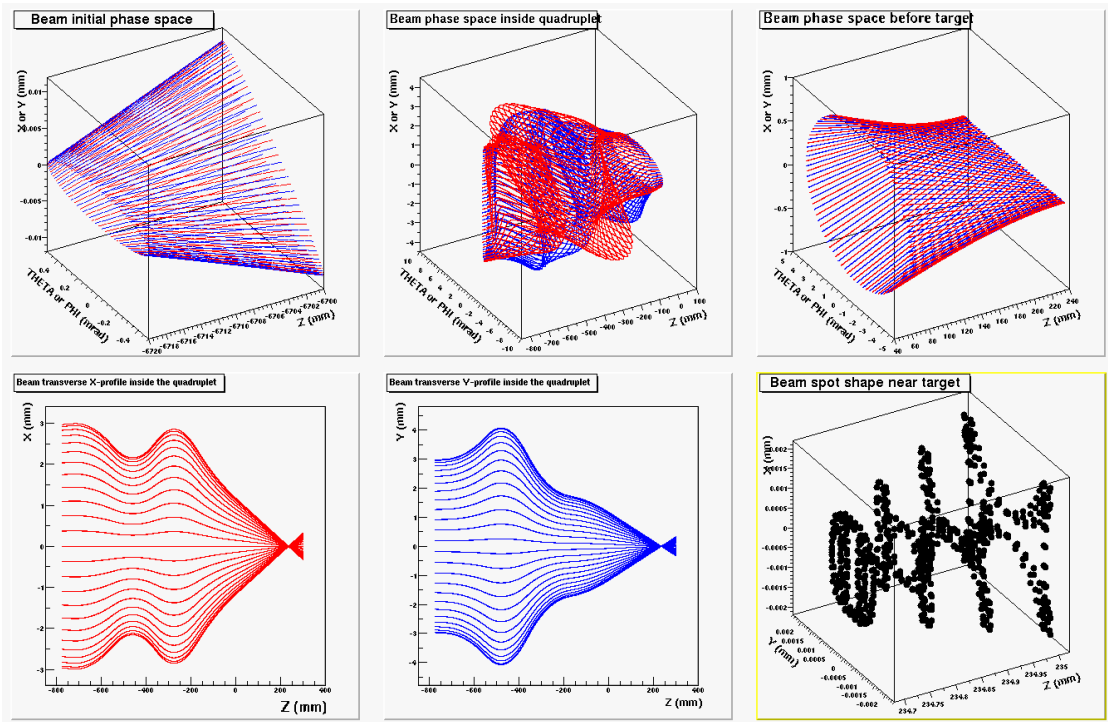

Fig. 2. The top plots show the evolution of the space phase of the beam external envelope - $(x, \theta)$ in red, $(y, \phi)$ in blue - from the collimator object (left), inside the quadruplet (middle) and before the target (right), along the beam propagation axis. The left and middle bottom plots show the beam profiles in the transverse X (red) and Y (blue) planes along the propagation axis. The right hand side plot shows the beam spot shape around the target position (235 mm). Note the typical distortions caused by the spherical aberrations of the quadruplet. Please see the electronic version for the figures in colour [8].

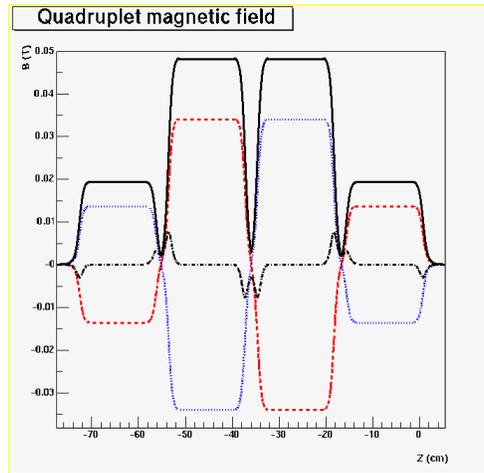

Fig. 3. Magnetic field components inside the quadruplet along the propagation axis ( $B_x$ in red dash, $B_y$ in blue dot-dot, $B_z$ in black dash-dot). The solid line shows the field total magnitude. The field has been calculated at the distance $r = 3\sqrt{2}$ mm from the propagation axis.

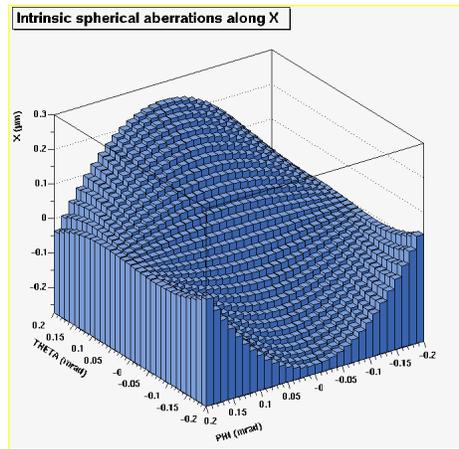

Fig. 4. Evolution of the X transverse position of the beam on target as a function of the incident angles $\theta, \phi$. A third order polynomial fit to this 2D curve allows the estimation of the system intrinsic spherical aberration coefficients $\langle x|\theta\rangle$, $\langle x|\theta\phi^2\rangle$ and $\langle x|\theta^3\rangle$.

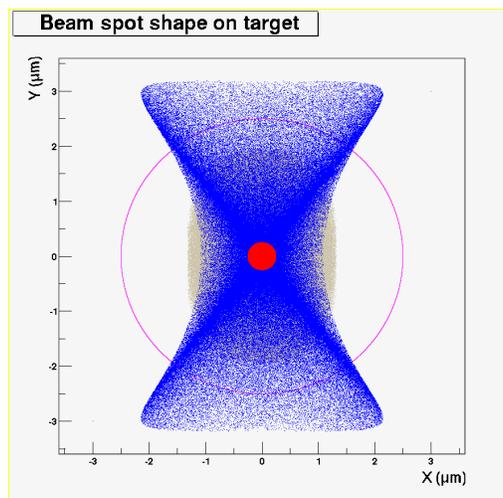

Fig. 5. Beam spot shape on target obtained assuming a uniform elliptical space phase distribution in both transverse planes. The central red spot shows the first order beam spot, with a diameter of 0.5 micrometer ; for comparison, the outer circle represents the 5 micrometer diameter object collimator. The grey area shows the second order chromatic aberrations contribution and the blue area (aisles) shows the third order spherical aberrations. Note the asymmetrical shape of the chromatic and spherical components, which does not appear on the first order collimator image.

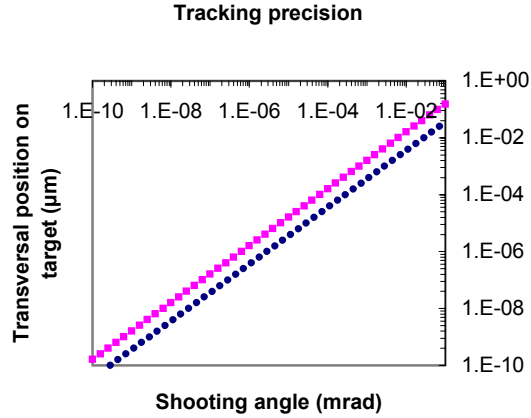

Fig. 6. Transverse displacement X (or Y) in the image plane as a function of shooting angles $\theta$ (or $\phi$) over a wide range of angles.

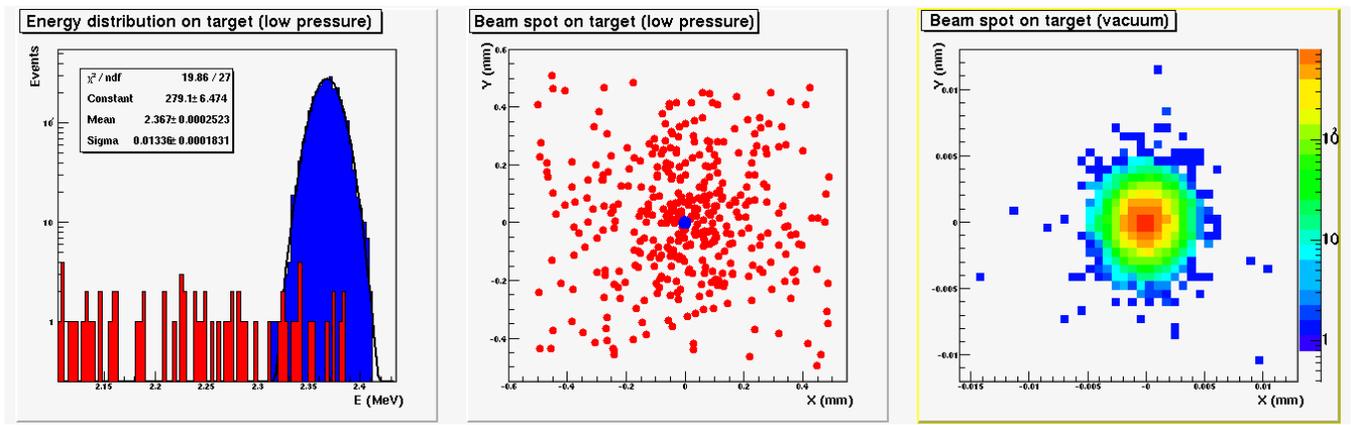

Fig. 7. The left plot shows in blue the alpha beam kinetic energy distribution on target, adjusted to a Gaussian distribution ($\langle T \rangle = 2.37 \pm 0.01$ MeV) and the red background shows alphas that have been scattered by the diaphragm edges before reaching the target. The middle plot shows the corresponding spatial distributions on target. The beam is spread inside a square of side slightly over 1 mm, corresponding to the surface of the $Si_3N_4$ window. Finally, the right plot shows the beam spatial distribution when the residual low pressure air inside the beam pipe has been replaced by vacuum. No diaphragm scattering has been observed in this case.

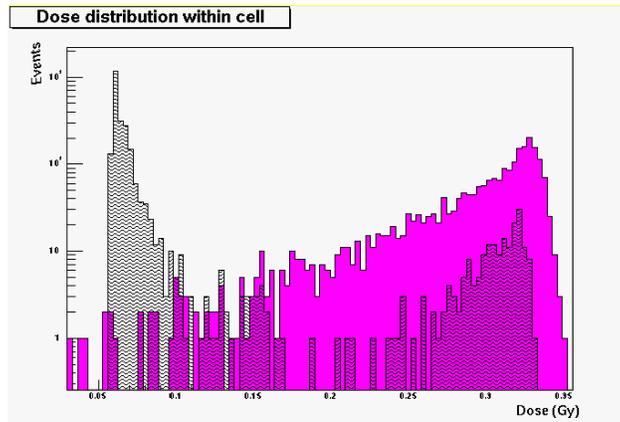

Fig. 8. Dose distributions in the HaCat cell geometrical model. The shaded distribution shows the dose distribution in the cytoplasm. The low dose part corresponds to particles crossing both the cytoplasm and the nucleus. The higher dose part is deposited by alpha particles that hit the cytoplasm only, without reaching the nucleus. The purple distribution shows the dose distribution in the nucleus.